# Geospatial Question Answering on Historical Maps Using Spatio-Temporal Knowledge Graphs and Large Language Models

Ziyi Liu
Institute of Cartography and
Geoinformation, ETH Zurich
Switzerland

Sidi Wu*
Institute of Cartography and
Geoinformation, ETH Zurich
Switzerland

Lorenz Hurni
Institute of Cartography and
Geoinformation, ETH Zurich
Switzerland

**Abstract**

Recent advances have enabled the extraction of vectorized features from digital historical maps. To fully leverage this information, however, the extracted features must be organized in a structured and meaningful way that supports efficient access and use. One promising approach is question answering (QA), which allows users—especially those unfamiliar with database query languages—to retrieve knowledge in a natural and intuitive manner. In this project, we developed a GeoQA system by integrating a spatio-temporal knowledge graph (KG) constructed from historical map data with large language models (LLMs). Specifically, we have defined the ontology to guide the construction of the spatio-temporal KG and investigated workflows of two different types of GeoQA —factual and descriptive. Additional data sources, such as historical map images and internet search results, are incorporated into our framework to provide extra context for descriptive GeoQA. Evaluation results demonstrate that the system can generate answers with a high delivery rate and a high semantic accuracy. To make the framework accessible, we further developed a web application that supports interactive querying and visualization.

## 1 Introduction

Historical maps are valuable resources that contain useful information about the Earth in the past and can be used to track object changes over time. They are relevant for various fields like urban planning, environmental science, ecology, and digital humanities. Recent advancements have enabled the accurate extraction of vectorized features from digital historical maps [1, 2, 5, 21–23], allowing for precise representations of map features. To effectively retrieve, interpret, and explore spatio-temporal information from historical maps, proper data structures, efficient querying strategies, as well as accessible and user-friendly system are necessary, the latter being especially important for non-domain experts.

Knowledge graphs (KGs) enable the representation of knowledge in a semantically enriched, formal, and structured way [7]. They are organized collections of entities and concepts linked via their relations with flexible structures, which can be used to store, link, and query data extracted [17, 18]. Sun et al. [19] proposed a unified framework for geospatial data ontology, which is essential to link and integrate data. For historical data, Shbita et al. [17] also use external sources like OpenStreetMap (OSM) to enrich the object information by matching instances at the same location. Additionally, text recognition has been applied on historical maps to enrich metadata and improve information search [4].

To query knowledge from geospatial KGs, the standard query language SPARQL and its geospatial extension GeoSPARQL are typically used. Initializing such queries and understanding raw queried results would pose barriers to non-experts. To overcome this challenge, researchers have worked on the development of Geographic Question Answering (GeoQA) systems that retrieve or generate answers in response to users' queries in natural language in the geospatial domain [11, 12]. While these methods are effective for handling well-structured and narrowly scoped queries, they suffer from limited flexibility due to their dependence on pre-defined templates. As a result, they struggle with more complex or descriptive questions that fall outside the scope of available templates. Moreover, they often fail to support correspondingly rich, explanatory answers, which are essential when addressing open-ended queries. Large language models (LLMs) have shown great performance in QA [14]. Nevertheless, LLMs occasionally fail to provide factual knowledge due to the lack of domain-specific knowledge. Some studies have investigated on embedding geospatial knowledge into LLMs by fine-tuning or training [8, 10, 26]. These require considerable training data and computational resources. To address this, other works leveraged pre-trained LLMs with no or few epochs of fine-tuning, using prompt engineering to guide geospatial tasks [6, 9, 13, 25].

In our project, we have proposed the ontology for constructing spatio-temporal KGs from vectorized geographic features extracted from historical maps. We integrated the constructed KG with LLMs to build a spatio-temporal GeoQA for both factual questions and more open-ended, descriptive questions. Our system not only parses users' questions in natural language but also generates answers in natural language, especially for descriptive questions, grounded on the facts from the constructed KGs. A web application is also developed to facilitate interactive querying and visualization.

## 2 Methods and implementation

### 2.1 Dataset

We test our approach on vectorized geographic features extracted from the Siegfried maps[1], a collection of topographic historical maps of Switzerland. We focus on one sample region, with four map sheets with a scale of 1:25,000 covering multiple historical years, including 1877, 1901, 1916, and 1930. The selected map sheets primarily cover or intersect with municipalities such as Aarberg, Bargen (BE), Seedorf (BE), and Radelfingen, in the canton of Bern.

---

*Corresponding author:sidiwu@ethz.ch

[1] https://www.swisstopo.admin.ch/en/siegfried-map

## 2.2 Ontology definition

The overall ontology is object-based. We define *geographic features* as the representations of geometry units[17]. For each feature, there are semantic properties and geospatial properties, as well as its spatial relations and temporal relations with other features. The semantic model of our ontology is illustrated in Figure 1.

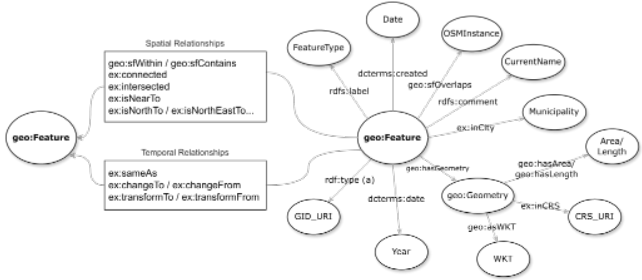

Figure 1: Semantic model for knowledge graph construction. Each feature has its own properties and relationships with other features.

## 2.3 Spatio-temporal KG construction

For feature properties, core properties such as feature type and year can be extracted directly. Additional properties like municipality and area or length are derived from feature geometries. Moreover, properties are further enriched by external data sources (i.e., OSM [17]), e.g., current names of features are obtained from their linked OSM instances. For feature relations, spatial relations include three aspects: topological, proximity, and cardinal directions[7, 19]. Temporal relations are defined based on their geometries. For example, we define "change" as changed to / from another feature of the same feature type and the next / previous time stamp; and "transform" as transform to / from another feature of a different feature type and the next / previous timestamp. Feature geometries are essential to geospatial KG construction. However, there could be uncertainty or misalignment in geometry extraction [11]. Relying strictly on feature geometries may lead to incorrect interpretations of their relationships. To eliminate the influence of positional inaccuracy, the most relevant relations are precomputed by defined metrics and buffer zones [16, 20], via executing GeoSPARQL queries defined. GraphDB[2] is used as our SPARQL or GeoSPARQL query endpoint.

## 2.4 Spatio-temporal question categorization

We propose a set of representative questions for evaluating our QA system. Taking current geospatial question sets and categorizations as reference [7, 11, 15], we generate question sets based on our KG and use GPT-4o to produce variation in phrasing. The question-content based categorization is visualized in Figure 2. Furthermore, based on answer-contents (i.e., clear or open-ended), these questions can be further categorized into factual questions and descriptive questions. Specifically, we generate 45 Yes/No questions and 45 numerical questions for factual questions, and 10 overview questions for descriptive questions.

[2]https://graphdb.ontotext.com/

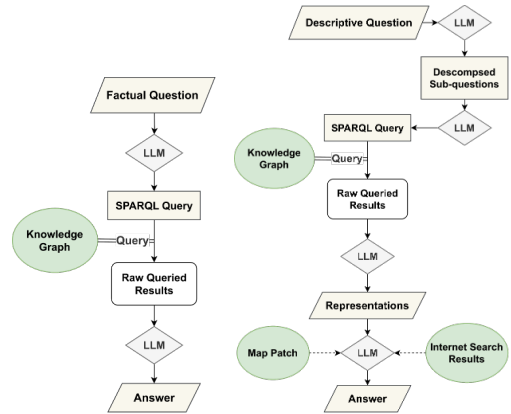

Figure 2: Geospatial question categorization with examples for our geospatial QA system. These questions can also be categorized into: factual questions including simple factoid and geo-analytical questions; and descriptive questions.

## 2.5 Geospatial QA workflow

Since factual questions and descriptive questions have different types of answers, we adopt two QA workflows to handle them separately, which are shown in Figure 3.

Figure 3: Factual (left) and descriptive (right) geospatial question answering workflows.

For factual questions, they are first converted into SPARQL queries using an LLM, guided by a structured prompt. The prompt includes analysis instructions (e.g., identifying place names with predefined choice lists), SPARQL modules with all fixed or optional properties and relations to avoid querying non-existent information, query generation constraints with rules to reduce ambiguity, and few-shot examples to improve performance. Before executing, the generated query is passed through a second LLM to perform a validation check. The queried results, along with the original question and query, are passed to the LLM to reason, rerank, and generate a natural language answer.

For descriptive questions, the system first decomposes them into sub-questions, each processed as a factual question. We further use a LLM to generate representations of queried results by converting them from JSON format into plain text. This is to reason over and extract question-related facts in advance, while also reducing context length due to LLM API input limits. Additional contextual information such as map image patches and internet search results are incorporated to generate the final answer. For internet search,

LangChain[3] LLM agent is used. Specifically, we use Tavily[4], which is a search engine built for LLM agents.

## 2.6 Web interface

A web application is developed to enable interactive SPARQL / GeoSPARQL querying, geospatial QA on factual and descriptive questions, and results visualization. The code is adapted from Shbita et al. [17] and the interface is shown in Figure 4.

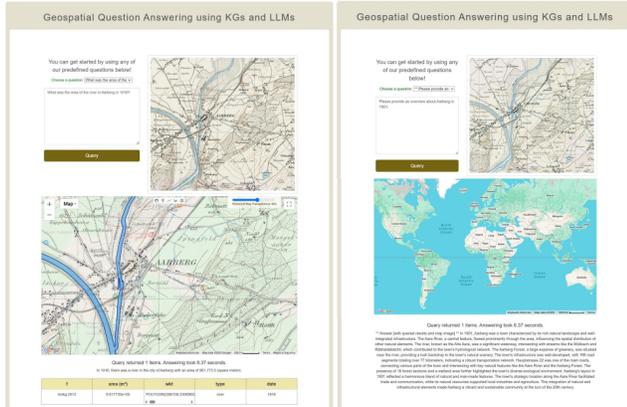

Figure 4: Web interface of our geospatial QA system, with examples of factual QA on the left and descriptive QA on the right.

## 2.7 Evaluation

**Factual QA** To evaluate our system's performance on factual QA, we adopt three metrics [24]:

- **Delivery Rate**: whether the system can deliver a result successfully.
- **Accuracy**: whether the delivered answer is correct.
- **SPARQL Accuracy**: whether the generated SPARQL is semantically correct.

We introduce a SPARQL semantic check via both GPT and manual review, to assess whether the generated query accurately interprets the question, particularly in cases of potential ambiguity such as municipality constraints, which should be clarified by rules in the prompt.

**Descriptive QA** Descriptive questions lack ground truth answers and their answering also incorporates external resources other than queried results, making comprehensive fact-checking difficult. Therefore, we focus on verifying KG-related facts. For example, given the question "What was Aarberg like in 1901?", part of the answer could be: *"The town was surrounded by 18 __forest__ sections, covering over 4 million square meters, and a single __wetland__ area of about 29,114 square meters, indicating a lush natural environment."*
We manually label KG-related features, i.e., forest and wetland, and their KG-related properties and relations (marked in orange). We then decompose the descriptive answers into several factual questions:

- *Were there 18 forest sections covering over 4 million square meters in Aarberg in 1901?*
- *Was there a single wetland area of about 29,114 square meters in Aarberg in 1901?*

We obtain the answers automatically, the factual QA workflow. We call this step fact checks. If the fact exists in the KG, the answer is "Yes". We define:

- **Fact Accuracy**: whether the KG-based fact stated is correct.

As shown in Table 1, the automatically generated SPARQL queries can be semantically incorrect, leading to erroneous results even when the factual questions are valid. To address this issue, we manually reviewed the cases with incorrect query results. In addition, we also evaluate the content quality. We first use perplexity[5] to measure the prediction confidence. Since we don't have ground truth to compare, we also refer to reference-free evaluation methods [3] to leverage the evaluation power of LLMs per see, by defining three extra criteria using DeepEval[6]. In total, we have:

- **Perplexity**: how confident the model is in predicting a sequence of words.
- **Relevance**: whether the provided answer addresses to the geospatial descriptive question.
- **Fluency**: the naturalness, fluency, and readability of the answer.
- **Informativeness**: whether the answer provides informative content relevant to the question.

## 3 Results

### 3.1 Factual QA

We evaluate our system's performance on factual QA and also compare the performance of different LLMs (e.g., Deepseek-Reasoner, GPT-4o, and Claude 3.7 Sonnet) as the SPARQL generator using the same prompt setting. Table 1 displays the results.

Table 1: Evaluation results of factual QA using different SPARQL generators. The arrow indicates the direction for better performance. We check the SPARQL accuracy via both GPT (auto) and manual review.

| SPARQL Generator | Delivery Rate↑ | Accuracy↑ | SPARQL Accuracy (auto)↑ | SPARQL Accuracy (manual)↑ |
|---|---|---|---|---|
| Deepseek-Reasoner | 0.98 | 0.88 | **0.72** | **0.84** |
| GPT-4o | **0.99** | **0.89** | 0.63 | 0.80 |
| Claude 3.7 Sonnet | 0.93 | 0.77 | 0.62 | 0.72 |

### 3.2 Descriptive QA

We compare different combinations of context sources for descriptive QA evaluation. GPT-4o is used as the SPARQL generator. The evaluation results are presented in Table 2.

## 4 Discussion

The results show that the system can successfully deliver correct results across different types of questions in most cases. Deepseek-Reasoner and GPT-4o achieved similar accuracy, superior to Claude 3.7 Sonnet. The accuracy of all three models is clearly higher with manual check compared to GPT check, because GPT is more rigid regarding logical or structural variations in SPARQL queries, which can also be valid. For descriptive QA, the KG-related fact accuracy with manual check for all context combinations is larger than 0.9, indicating that relevant knowledge is correctly preserved. However, with more contextual sources, the system tends to generate answers containing less KG-related facts, especially for the configuration with internet search. This also explains its high accuracy. Integrating the map image achieves the best content quality, leveraging the power of GPT for visual interpretation. However, there are also limitations. While pre-computing feature relations helps address geometric uncertainty, it would reduce QA's dynamic reasoning flexibility. When generating SPARQL queries, ambiguities in question interpretation cannot be fully resolved by defining rules in the prompt. Moreover, the evaluation process, especially for descriptive QA, still requires manual check.

---

[3]https://www.langchain.com/
[4]https://tavily.com/
[5]https://huggingface.co/spaces/evaluate-metric/perplexity
[6]https://github.com/confident-ai/deepeval

Table 2: Evaluation results of descriptive QA using different combinations of context sources. We measure the fact accuracy and number of decomposed factual questions. The arrow indicates the direction for better performance.

| Context Sources | Fact accuracy (Auto)↑ | Fact Accuracy (Manual)↑ | Factual questions | Perplexity↓ | Relevance↑ | Fluency↑ | Informativeness↑ |
|---|---|---|---|---|---|---|---|
| KG | 0.67 | 0.91 | 9.4 | 30.89 | 0.93 | 0.77 | 0.85 |
| KG+Map | 0.77 | 0.93 | 7.5 | 29.80 | 0.97 | 0.87 | 0.89 |
| KG+Map+Semantic Map | 0.75 | 0.94 | 6.7 | **27.44** | **0.99** | **0.94** | **0.97** |
| KG+Map+Internet Search | **0.95** | **1.00** | 2.0 | 29.15 | 0.96 | 0.86 | 0.88 |

## 5 Conclusion and outlook

We have developed a GeoQA system on historical maps for factual and descriptive questions by constructing a spatio-temporal KG and integrating it with LLMs. Results show that our system can provide fact-grounded answers in natural language. To further enhance the system, more intelligent LLM agents can be used, while the combination of extra contextual sources with KG should be carefully investigated. Text recognition could be used to enrich feature metadata. More automated evaluation regarding the descriptive questions should be investigated. Furthermore, the usefulness and reliability of the answers can be potentially evaluated through user studies.